\documentclass[prd,aps,preprint,tightenlines,nofootinbib,superscriptaddress]{revtex4-1}
\usepackage{mathrsfs}
\usepackage{amsfonts}
\usepackage{amsmath}
\usepackage{slashed}
\usepackage{array}
\usepackage{verbatim}
\usepackage{epsfig}
\usepackage{graphicx}
\usepackage{color}

\newcommand{\beq}{\begin{eqnarray}}
\newcommand{\eeq}{\end{eqnarray}}

\begin{document}

\title{Hunting the Gluon Orbital Angular Momentum at the Electron-Ion Collider}

\author{Xiangdong Ji}
\affiliation{Maryland Center for Fundamental Physics, Department of Physics, University of Maryland, College
Park, Maryland 20742, USA}
\affiliation{INPAC, Department of Physics and Astronomy, Shanghai Jiao Tong University, Shanghai, 200240, P. R. China}

\author{Feng Yuan}
\affiliation{Nuclear Science Division, Lawrence Berkeley National
Laboratory, Berkeley, CA 94720, USA}

\author{Yong Zhao}
\affiliation{Maryland Center for Fundamental Physics, Department of Physics, University of Maryland, College
Park, Maryland 20742, USA}
\affiliation{Nuclear Science Division, Lawrence Berkeley National
Laboratory, Berkeley, CA 94720, USA}

\begin{abstract}
Applying the connection between the parton Wigner distribution and orbital 
angular momentum (OAM), we investigate the probe of the gluon OAM in hard
scattering processes at the planned electron-ion collider. We show that the 
single longitudinal target-spin asymmetry in the hard diffractive dijet production
is very sensitive to the gluon OAM distribution. The associated spin
asymmetry leads to a characteristic azimuthal angular correlation 
of $\sin(\phi_q -\phi_\Delta)$, where $\phi_\Delta$ and 
$\phi_q$ are the azimuthal angles of the proton momentum transfer and the 
relative transverse momentum between the quark-antiquark pair. This study will enable 
a first measurement of the gluon OAM in the proton spin sum rule.
\end{abstract}

\maketitle

\section{Introduction}

In the past three decades, we have witnessed significant advances in the 
understanding of high-energy hadron structure. Great progress has been made
in measuring the partonic content of proton spin, as results from SLAC, 
CERN, DESY, JLab and RHIC have nailed the quark spin contribution to about 
$30\%$~\cite{deFlorian:2009vb,Nocera:2014gqa}. 
More recently, RHIC experiments have revealed that the 
gluon polarization contributes about $40\%$ within the kinematic range of 
$0.05\le x\le0.2$~\cite{deFlorian:2014yva}, which is an important part of
the proton spin sum rule~\cite{Jaffe:1989jz}. 
With the completion of JLab 12 GeV upgrade and implementation of the Electron Ion 
Collider (EIC), the proton spin structure will be studied to an unprecedented extent 
with higher precision.  Among them, the major focus will be the gluon helicity
distribution at smaller $x$, and in particular, the orbital angular momenta (OAM) 
from the quarks and gluons~\cite{Accardi:2012qut,Boer:2011fh}. 
The latter play important roles in the partonic structure in nucleon, 
not only for the proton spin sum rule, but also 
for the novel phenomena in various high energy scattering processes. It has 
been shown in~\cite{Ji:1996ek} that the total angular momentum 
contributions from the quarks and gluons
can be studied through the associated generalized parton distributions 
(GPDs)~\cite{Mueller:1998fv,Ji:1996nm,Radyushkin:1997ki} 
measured in the hard exclusive processes,
such as the Deeply Virtual Compton Scattering (DVCS)~\cite{Ji:1996ek,Ji:1996nm}. 
By subtracting the helicity contributions, we will be able to obtain the 
corresponding OAM contributions from the quarks and gluons.

Recent developments have also unveiled the close connection between
the parton OAM and the associated quantum phase space distributions,
the so-called Wigner distribution functions~\cite{Belitsky:2003nz,Lorce:2011kd,Hatta:2011ku,
Lorce:2011ni,Ji:2012sj},
\begin{equation}
L_{q,g}(x) = \epsilon_\perp^{\alpha\beta} \left.{\partial \over i\partial \Delta_\perp^\alpha}\right|_{\Delta=0}\int d^2k_\perp \ k^\beta_\perp f_{q,g}(x, \xi, k_\perp,\Delta_\perp)\ ,
\end{equation}
where $f_{q,g}$ represent the quark/gluon Wigner
distributions in a longitudinal polarized nucleon, 
and $\epsilon^{\alpha\beta}_\perp$ represents 
2-dimensional Levi-Civita symbol. 
We focus on the gluon Wigner distribution with light-cone
gauge links and the corresponding OAM belongs to the Jaffe-Manohar
spin sum rule~\cite{Ji:2012ba,Hatta:2012cs}.
The Wigner distributions are also referred to as the generalized transverse
momentum dependent parton distributions~\cite{Meissner:2009ww}. 
This opens a new window to directly access the parton OAM
contributions to the proton spin. The goal of this paper is to show that
indeed that we can probe the gluon OAM distribution through the hard 
scattering processes in high energy lepton-nucleon collisions, in particular,
at the EIC.

\begin{figure}[tbp]
\centering
\includegraphics[width=0.8\textwidth]{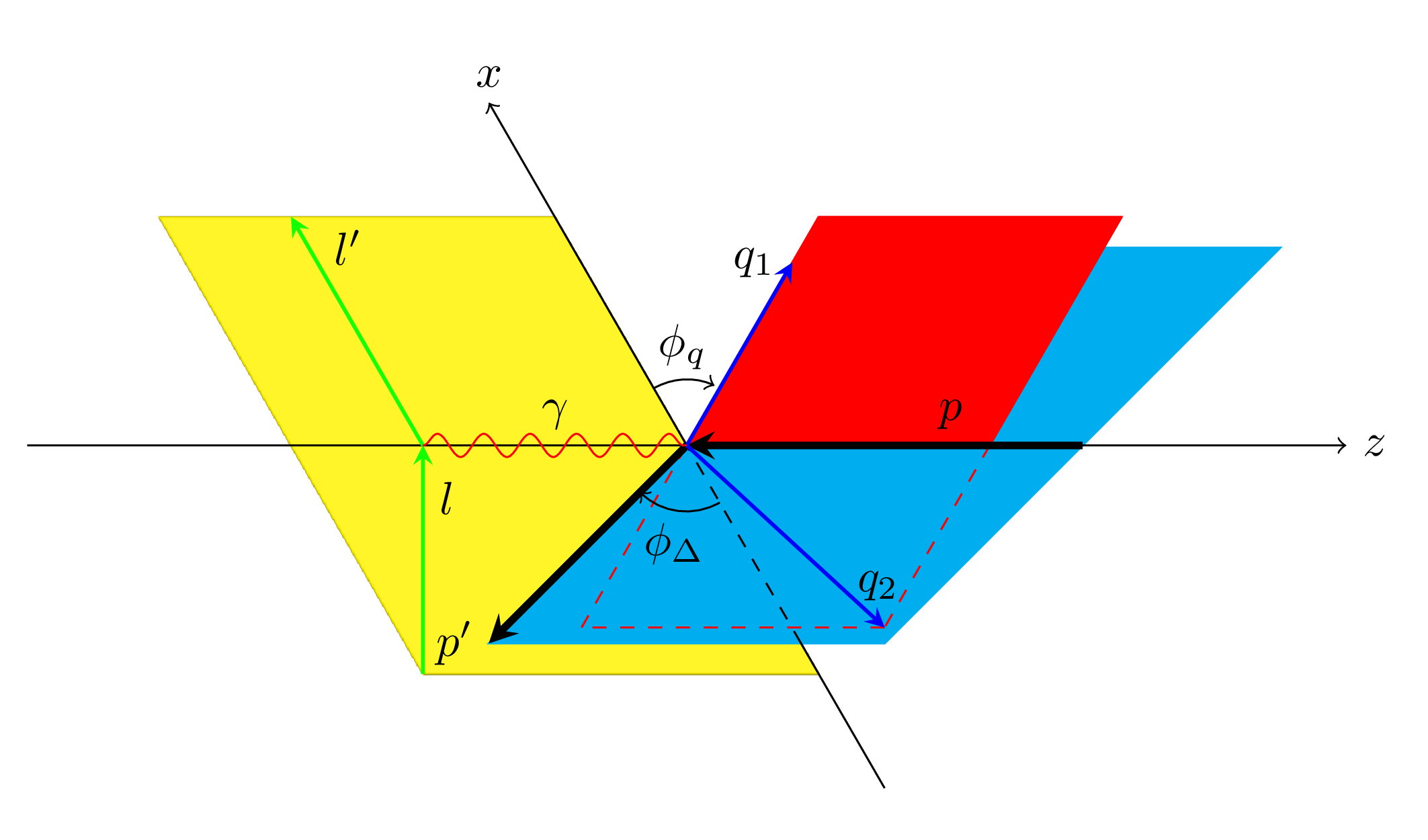}
\caption{Hard exclusive dijet production in deep inelastic scattering to
probe the gluon orbital angular momentum.}
\label{dijet}
\end{figure}

We take the example of the single longitudinal target-spin asymmetries
in hard exclusive dijet production in lepton-nucleon collisions~\cite{Braun:2005rg},
\begin{equation}
\ell+p\to \ell' +q_1+q_2+p' \ ,
\end{equation}
where the incoming and outgoing leptons have momenta $l$ and $l'$, proton
momenta with $p$ and $p'$, and the final state two jets with momenta $q_1$ and
$q_2$, as illustrated in Fig.~\ref{dijet}. In high energy experiments
at the EIC, the process of (2) is dominated by the gluon distribution
from the target nucleon, and in particular, the differential cross section will
depend on the gluon Wigner distribution~\cite{Hatta:2016dxp}. 
Because of the relation of Eq.~(1), one expects that the single longitudinal target-spin
asymmetry of this process will be an ideal probe to the gluon OAM. 
To show this explicitly, we perform our calculations in a general collinear
factorization framework, where the gluon OAM distribution 
enters at the twist-three level. The spin dependent differential cross
section has a characteristic azimuthal angular dependence of 
$\sin(\phi_\Delta -\phi_q)$ where $\phi_\Delta$ and $\phi_q$ are
the azimuthal angles of the proton momentum transfer and the
relative transverse momentum between the quark pair as shown
in Fig.~\ref{dijet}. With a hermetic detector designed for the EIC,
this observable can be well studied in the future, and 
will help us to finalize the proton spin sum rule, the ultimate goal for 
hadron physics in past decades.

We notice that there have been other proposals to measure the parton 
OAMs~\cite{Courtoy:2013oaa,Courtoy:2014bea}. 
Our approach and observables are different from theirs. In particular,
we focus on the hard scattering processes which can be well studied
at the planed EIC.
The rest of this paper is organized as follows. In Sec. II, we derive
the single longitudinal target-spin asymmetry in hard exclusive dijet 
production in lepton-nucleon collisions. We take the leading contribution from the gluon OAM
distribution in the nucleon. We summarize our results and comment
on further developments in Sec. III.

\section{Gluon OAM Contribution to the Single Spin Asymmetries}
\label{kinematics}

\begin{figure}[tbp]
\centering
\includegraphics[width=7cm]{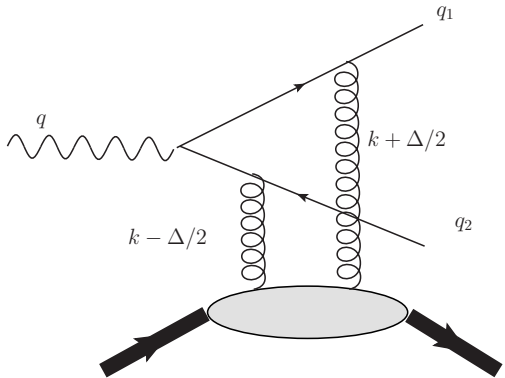}
\caption{Generic Feynman diagram to evaluate the single longitudinal spin asymmetry in 
the hard exclusive dijet production in deep inelastic lepton nucleon 
scattering processes. All possible gluon attachment has been included in our 
calculations.}
\label{feynman}
\end{figure}

The differential cross section of process (2) can be calculated through the
lepton tensor and hadronic tensor,
\begin{eqnarray}
|{\cal M}|^2=L_{\mu\nu}H^{\mu\nu}\ ,
\end{eqnarray}
where the lepton tensor takes a simple form of $L_{\mu\nu}=2(l_\mu l'_\nu+l_\nu l'_\mu-g_{\mu\nu}l\cdot l')$ due
to the fact that the incoming lepton is unpolarized. The main task of our calculation
is to evaluate the hadronic tensor, which comes from the Feynman diagrams 
illustrated in Fig.~\ref{feynman}. We adopt the 
usual kinematics: the incoming photon with momentum
$q = l - l'$, $q^2 =-Q^2$, $x_{Bj}= {Q^2/( 2q\cdot p)}$, $y= {q\cdot p /(l\cdot p)}$. 
The quark and antiquark momenta are further parameterized by their longitudinal
momentum fractions $z$ and $\bar z=1-z$ as well as their transverse momenta 
$q_\perp-\Delta_\perp/2$ and $-q_\perp-\Delta_\perp/2$.
In addition, for the exclusive processes, we have the following kinematics: 
$\Delta =p'-p$, $P = (p+p')/2$, $t=\Delta^2$,  $(q+p)^2 = W^2$, 
$(q-\Delta)^2=(q_1+q_2)^2 = M^2$, and the skewness parameter is defined as 
$\xi = {(p^+ - p'^+)/( p^+ + p'^+)}$ with $p^\pm=(p^0\pm p^z)/\sqrt{2}$, where
$q$ and $p$ are chosen to be along the $z$ axis. As shown in Fig.~\ref{dijet}, the lepton plane
is set as the $x-z$ plane. The quark pair are in one plane with azimuthal angle $\phi_q$
respect to the lepton plane, whereas the recoiled proton is in another plane with momentum 
transfer $\vec{\Delta}_\perp$ and azimuthal angle $\phi_\Delta$.
The spin-average cross section for this process has been calculated in 
Ref.~\cite{Braun:2005rg}. In the following, we will compute the single longitudinal target-spin 
asymmetry. We will show how this asymmetry can be related to the gluon OAM contributions.

Generically, the single longitudinal spin asymmetry in the above process 
can be evaluated following the usual collinear expansion at the next-to-leading 
power. We write the scattering amplitude, depicted in Fig.~\ref{feynman}, as
\begin{eqnarray}
i\mathcal{A}_f &\propto & \int dxd^2k_\perp
\mathcal{H}(x,\xi, q_\perp,k_\perp,\Delta_\perp)\ xf^g(x, \xi,k_\perp, \Delta_\perp)\ ,
\end{eqnarray}
where $q_\perp$ is the jet transverse momentum defined above, and $k_\perp$ is the 
gluon transverse momentum entering the hard partonic part of Fig.~\ref{feynman}. 
In this calculation, $q_\perp$ is the same order of $Q$, while the nucleon recoil 
momentum $\Delta_\perp$ is much smaller than $Q$. 
In the twist analysis, we expand the scattering amplitude in terms of $k_\perp/q_\perp$
(or $k_\perp/Q$), 
\begin{equation}
{\cal H}(x,\xi,q_\perp,k_\perp,\Delta_\perp)={\cal H}^{(0)}(x,\xi,q_\perp,0,\Delta_\perp)+k_\perp^\alpha
\frac{\partial}{\partial k_\perp^\alpha}{\cal H}(x,\xi,q_\perp,0,\Delta_\perp) +\cdots \ . \label{e5}
\end{equation}
For the spin-average cross section, we take the zero-th order expansion of $k_\perp$. As a
result, $k_\perp$ is integrated out for the gluon Wigner distribution,
\begin{equation}
\int d^2k_\perp xf^g(x,\xi,k_\perp,\Delta_\perp)=F_g(x,\xi,\Delta_\perp)\ ,
\end{equation}
where $F_g$ is the spin-average gluon GPD. The scattering amplitude can be written as
\begin{equation}
i\mathcal{A}_f^{(0)} \propto  \int dx
\mathcal{H}^{(0)}(x,\xi, q_\perp,0,0)\ xF_g(x, \xi, \Delta_\perp)\ .
\end{equation}
Because $\Delta_\perp\ll q_\perp$, we have also taken $\Delta_\perp=0$ in the hard
partonic part. This will enter into the spin-average cross section contribution, e.g.,
Eq.~(\ref{unp}) below.


On the other hand, the single longitudinal target-spin asymmetry comes from the next-to-leading
power expansion of Eq.~(\ref{e5}). Because of the nontrivial correlation between $k_\perp$ and $\Delta_\perp$
in the gluon Wigner distribution due to the gluon orbital motion, this contribution will lead to a 
novel correlation between $q_\perp$ and $\Delta_\perp$ as mentioned in Introduction,
\begin{eqnarray}
\int d^2k_\perp (\vec{q}_\perp\cdot \vec{k}_\perp) xf^g(x,\xi,k_\perp,\Delta_\perp)=-iS^+(\vec{q}_\perp\times \vec{\Delta}_\perp) xL_g(x,\xi,\Delta_\perp)+\cdots\ ,
\end{eqnarray}
where we have only kept the spin-dependent matrix element in the above equation and
$S^+$ represents the longitudinal spin, and we have taken the leading contribution
in terms of $(\vec{q}_\perp\cdot \vec{k}_\perp)$ in ${\cal H}$. We refer the above $L_g(x,\xi,\Delta_\perp)$
as the gluon OAM distribution, from which we shall be able to obtain the gluon OAM
contribution to the proton spin from Eq.~(1).  
According to this result, we only need to measure how the single target-spin 
asymmetry modulates with $\sin(\phi_q-\phi_\Delta)$---
which comes from ($\vec{q}_\perp \times \vec{\Delta}_\perp)$---to extract 
the gluon OAM density. 

The detailed derivations will be presented in a separate publication. Here, we present the main
results and demonstrate the sensitivity of the spin asymmetries on 
the gluon OAM distribution. For the spin-average cross section, we have the following expression~\cite{Braun:2005rg},
\begin{eqnarray}
{d\sigma \over dy dQ^2 d\Omega} &=& \sigma_0\left[ (1-y)|{A}_L|^2+ \frac{1+(1-y)^2}{2}|{A}_T|^2\right]
\ , \label{unp}
\end{eqnarray}
where $d\Omega$ represents the final hadronic states phase space: $d\Omega=dzdq_\perp^2 d\Delta_\perp^2 d\phi_{q\Delta}$.
$\sigma_0$ is defined as
\begin{equation}
\sigma_0=\frac{\alpha_{em}^2\alpha_s^2e_q^2}{16\pi^2Q^2yN_c}\frac{4\xi^2 z\bar z}{(1-\xi^2)(\vec{q}_\perp^2+\mu^2)^3}\ ,
\end{equation}
where $\mu^2=z\bar z Q^2$,
and we have only kept the 
azimuthal angular symmetric terms in the above result and $\phi_{q\Delta}=\phi_q-\phi_\Delta$.
The contributions from the transverse and longitudinal photons are:
$|A_L|^2= 4 \beta\left|\mathcal{F}_g + 4\xi^2\bar{\beta} \mathcal{F}'_g\right|^2 $,
$|A_T|^2=\bar{\beta} \left(1/(z\bar{z})- 2\right)\left| \mathcal{F}_g + 2\xi^2(1-2\beta)\mathcal{F}'_g\right|^2 $, where 
$\beta=\mu^2/(\mu^2+\vec{q}_\perp^2)$. We have defined the following generalized Compton form factors,
\begin{eqnarray}
\mathcal{F}_g(\xi,t) &=&\int dx {1\over (x+\xi-i\varepsilon)(x-\xi+i\varepsilon)} F_g(x,\xi,t) \ ,\nonumber\\
\mathcal{F}'_g(\xi,t) &=&\int dx {1\over (x+\xi-i\varepsilon)^2(x-\xi+i\varepsilon)^2} F_g(x,\xi,t) \ .
\end{eqnarray}
Following the above procedure, we derive the longitudinal target-spin dependent differential cross section, 
\begin{eqnarray}
{d\Delta\sigma\over dy dQ^2 d\Omega} &=& \sigma_0\lambda_p
{2(\bar{z}-z)(\vec{q}_\perp\times \vec{\Delta}_\perp)\over \vec{q}_\perp^2 + \mu^2}\left[ (1-y){A}_{fL}+ \frac{1+(1-y)^2}{2}{A}_{fT}\right]
\ , \label{e12}
\end{eqnarray}
where $\Delta\sigma=(\sigma (S^+)-\sigma(-S^+))/2$ and $\lambda_p$ 
represents the longitudinal polarization for the incoming nucleon.
The spin-dependence comes from the interferences between the leading-twist and 
and twist-three amplitudes,  
\begin{eqnarray}
A_{fL}&=&16\beta\ \mbox{Im}\left(\left[ \mathcal{F}^*_g +4\xi^2\bar{\beta} \mathcal{F}'^*_g\right]\left[ \mathcal{L}_g +8\xi^2\bar{\beta} \mathcal{L}'_g\right]\right)\ ,\nonumber\\
A_{fT}&=& 2\ \mbox{Im}\left( \left[ \mathcal{F}^*_g + 2\xi^2(1-2\beta)\mathcal{F}'^*_g \right]\left[ {\cal L}_g+2\bar{\beta}
\left(\frac{1}{z\bar z}-2\right)
\left(\mathcal{L}_g + 4\xi^2(1-2\beta)\mathcal{L}'_g\right)\right]\right)\ ,
\end{eqnarray}
where, again, we have defined the following Compton form factors to simplify the final results,
\begin{eqnarray}
\mathcal{L}_g(\xi,t) &=&\int dx {x\xi\over (x+\xi-i\varepsilon)^2(x-\xi+i\varepsilon)^2}\ xL_g(x,\xi,t) \ ,\nonumber\\
\mathcal{L}'_g(\xi,t) &=&\int dx {x\xi\over (x+\xi-i\varepsilon)^3(x-\xi+i\varepsilon)^3}\ xL_g(x,\xi,t) \ .
\end{eqnarray}
The above equations are the main results of our paper. Clearly, because of the pre-factor of Eq.~(\ref{e12}),
we find that the spin asymmetry is a power correction in this process, which is consistent with our analysis.
In addition, it is proportional to $(\vec{q}_\perp\times \vec{\Delta}_\perp)$, so that it has the characteristic 
azimuthal angular correlation $\sin (\phi_q-\phi_\Delta)$.  

Similar calculations can be performed for the quark channel contributions, which may 
play important roles in the large-$x$ region. We will leave that for a future publication. 

\section{Discussion and Summary}

As shown above, the gluon OAM contribution to the single longitudinal target-spin
asymmetry has the novel azimuthal angular correlation of $\sin(\phi_q-\phi_\Delta)$. 
Experimentally, we have to identify the azimuthal angles of both $\vec{q}_\perp$ and 
$\vec{\Delta}_\perp$. In particular, it is challenging to precisely measure $\Delta_\perp$, because 
majority of the events will have small momentum transfer. Fortunately, the current design for
the EIC detector will have excellent coverage to study the $\Delta_\perp$ distribution in the
hard diffractive processes, including the proposed measurement of this paper, especially with
the Roman Pot device along the beam line of the EIC. With the measurements of these two
azimuthal angels, we can form the spin asymmetry,
\begin{eqnarray}
A_{\sin(\phi_q-\phi_\Delta)} &=& \int d\phi_q d\phi_\Delta\ {d\sigma_\uparrow - d\sigma_\downarrow\over d\phi_q d\phi_\Delta}\sin(\phi_q - \phi_\Delta)\left/ \int d\phi_q d\phi_\Delta\ {d\sigma_\uparrow + d\sigma_\downarrow\over d\phi_q d\phi_\Delta}\right.\ .
\label{obs}
\end{eqnarray}
From the results in the last section, we know that the above asymmetry will be sensitive to the 
gluon OAM distribution, and has the following kinematic dependence, schematically, 
\begin{equation}
A_{\sin(\phi_q-\phi_\Delta)}\propto {(\bar{z}-z)|\vec{q}_\perp| |\vec{\Delta}_\perp|\over \vec{q}_\perp^2 + \mu^2}\ ,
\label{asymmetry}
\end{equation}
where again, it is a twist-three effect. The size of the asymmetry, of course, will 
depend on how large the gluon OAM is. We are planing to have model
predictions for the typical kinematics at the EIC, and hope this will
lead to the first measurement of gluon OAM in the future.

In summary, we have calculated the differential cross section for the hard exclusive 
electro-production of quark-antiquark pair. The leading contribution to the single 
target-spin asymmetry is at order $1/Q$, and crucially depends on the gluon OAM.
In the kinematics covered by the EIC, this observable can be well studied, 
which will provide important information on the gluon OAM distribution in return. 

Being a higher-twist effect, the asymmetry defined in Eq.~(\ref{obs}) will
have contributions from the twist-three multi-parton GPDs, in particular,
those associated with three-gluon correlations~\cite{Ji:2012ba,Hatta:2012cs}. 
For a complete evaluation of the single longitudinal target-spin asymmetries,
we need to include these terms as well. Extension to other processes, such as the 
DVCS, will be also important to build a systematic framework to investigate the 
comprehensive tomography of partons inside the nucleons. We will address 
these issues in the future.

In this paper, we focus on moderate $x$ range of the gluon OAM
distribution. In the large-$x$ region, we will also have quark-channel 
contributions, which can be used to probe the quark OAM.
 At small-$x$, on the other hand, we would expect the dipole framework 
is more appropriate to compute the process (2)~\cite{Hatta:2016dxp}. However, the
spin asymmetry is much more involved and nontrivial, which will be addressed
in an accompanying paper by Hatta, Nakagawa, and the two of us~\cite{Hatta1}.

{\it Acknowledgments:}
We thank 	Elke-Caroline Aschenauer and Ernst Sichtermann for useful 
discussions on the experiment perspectives of the process studied in this paper. 
We also thank Yoshitaka Hatta and Yuya Nakagawa for comments and collaborations
on the related subject~\cite{Hatta1}.
This work was partially supported by the Laboratory Directed Research and Development 
Program of Lawrence Berkeley National Laboratory, the U.S. Department of Energy Office of 
Science, Office of Nuclear Physics under Award Number DE-FG02-93ER-40762 
and DE-AC02-05CH11231, and a grant (No. 11DZ2260700) from the Office of
Science and Technology in Shanghai Municipal Government, and also by 
grants from the National Science Foundation of China (No. 11175114, No. 11405104).

\end{document}